\documentclass[preprint,showpacs,preprintnumbers,amsmath,amssymb]{revtex4}
\usepackage{amssymb}
\usepackage{amsmath}
\usepackage{graphicx}
\usepackage{subfigure}
\usepackage{epsfig}
\usepackage{epstopdf}
\usepackage[usenames,dvipsnames]{color}
\usepackage{dcolumn}
\usepackage{bm}

\begin{document}

\title{Localization and quasilocalization of spin--$1/2$ fermion field
          on two-field thick braneworld}

\author{Heng Guo}\email[Corresponding author:~]{hguo@xidian.edu.cn}
\affiliation{School of Physics and Optoelectronic Engineering, Xidian University, Xi'an 710071, China}

\author{Qun--Ying Xie}\email[]{xieqy@lzu.edu.cn}
\affiliation{School of Information Science and Engineering, Lanzhou University,
    Lanzhou 730000, China}

\author{Chun--E, Fu}\email[]{fuche13@mail.xjtu.edu.cn}
\affiliation{School of Science, Xi'an Jiaotong University, Xi'an 710049, China}


\begin{abstract}
Localization of a spin--$1/2$ fermion on the braneworld is an important and interesting problem. It is well known that a five--dimensional free massless fermion $\Psi$ minimally coupled to gravity cannot be localized on the Randall-Sundrum braneworld. In order to trap such a fermion, the coupling between the fermion and bulk scalar fields should be introduced. In this paper, localization and quasilocalization of a bulk fermion on the thick braneworld generated by two scalar fields (a kink scalar $\phi$ and a dilaton scalar $\pi$) are investigated. Two types of couplings between the fermion and two scalars are considered. One coupling is the usual Yukawa coupling $-\eta\bar{\Psi}\phi\Psi$ between the fermion and kink scalar, another one is $\lambda\bar{\Psi}\Gamma^{M}\partial_{M}\pi\gamma^{5}\Psi$ between the fermion and dilaton scalar. The left--chiral fermion zero mode can be localized on the brane, and both the left-- and right--chiral fermion massive Kaluza--Klein modes may be localized or quasilocalized. Hence the four--dimensional massless left--chiral fermion and massive Dirac fermions, whose lifetime is infinite or finite, can be obtained on the brane.
\end{abstract}


\maketitle

\section{\label{sec:intro}introduction}

The braneworld scenarios that our observed four--dimensional Universe can be considered as a brane embedded in a higher--dimensional space--time, can supply new insights for solving the gauge hierarchy problem \cite{gog,ADD} and the cosmological constant problem \cite{RubakovPLB1983136,Randjbar-DaemiPLB1986,CosmConst}. The effective four--dimensional gravity could be recovered even in the case of non--compact extra dimensions in the Randall-Sundrum (RS) braneworld model \cite{Randall-Sundrum}, where singularities are present at the position of the branes. The smooth braneworld solutions (thick brane scenario or domain wall scenario) are generally based on gravity coupled to one or several bulk scalar fields. For some comprehensive reviews about thick branes, please see Refs. \cite{0812.1092,0904.1775,0907.3074,1003.1698,brane_book,1004.3962}.

In braneworld scenarios, an interesting and important problem is whether various bulk fields can be confined to the brane by a natural mechanism. Generally, a free massless scalar field \cite{BajcPLB2000} can be localized on branes of different types. A free vector field can be localized on the Randall-Sundrum brane in some higher--dimensional cases \cite{OdaPLB2000113} or on some thick de-Sitter (dS) branes and Weyl thick branes \cite{Liu0708,PRDGuo13}.

Since the particles in our world are described by fermions, localization of a bulk fermion is very important for any braneworld model  \cite{BajcPLB2000,OdaPLB2000113,Liu0708,PRDGuo13,NonLocalizedFermion,NonLocalFerm,
IchinosePRD2002,Ringeval,GherghettaPRL2000,Neupane,RandjbarPLB2000,KoleyCQG2005,
DubovskyPRD2000,0803.1458,0901.3543,Liu0907.0910,Koley2009,LiuJCAP2009,Liu0803,
PRD83:045002,CQG28,PRD105010,EPJC71,TMP175,CQG31,liuPRD89,LiuJHEP2007,0812.2638,
skyrmionbrane,PLBFermistring}. In general, if one does not introduce the coupling between the fermion and scalars which generate the thick braneworlds, a bulk fermion does not have a normalizable zero mode in five dimensions. So in order to localize a bulk fermion, the scalar--fermion coupling should be introduced. Usually, the Yukawa coupling $-\eta\bar{\Psi}\phi\Psi$ is a natural choice when the scalar has the configuration of a kink (i.e., $\phi$ is an odd function of the extra dimensional coordinate). However, if the background scalar has the shape of lump (or an even function of the extra dimensional coordinate, see Refs. \cite{two-scalarbrane1,KRthickbrane,twofieldbrane,deformedBrane,yangPRD86,yangIJMPA}), the Yukawa coupling cannot keep the $Z_{2}$ reflection symmetry of Lagrange and hence this localization mechanism does not work anymore. Recently, Liu et al. presented a new localization mechanism of a bulk fermion on braneworlds generated by such a scalar with even parity \cite{liuPRD89}. The scalar--fermion coupling is given by $\lambda\bar{\Psi}\Gamma^{M}\partial_{M}\pi\gamma^{5}\Psi$, which ensures not only the $Z_2$ reflection symmetry of the Lagrange but also localization of the left-- or right--chiral fermion zero mode on the branes.

In this paper, we will explore localization of a spin--$1/2$ fermion $\Psi$ on the Minkowski brane generated by two scalar fields (a kink scalar field $\phi$ with odd parity and a dilaton scalar field $\pi$ with even parity). We introduce both two types of couplings between the fermion and scalars: one is usual Yukawa coupling $-\eta\bar{\Psi}\phi\Psi$ between the fermion and kink scalar, anther is the coupling $\lambda\bar{\Psi}\Gamma^{M}\partial_{M}\pi\gamma^{5}\Psi$ between the fermion and dilaton scalar. If only introducing the Yukawa coupling, the effective potentials of the left-- and right--chiral fermion Kaluza--Klein (KK) modes in the corresponding Schr\"{o}dinger equations are modified volcano--type potentials. And the left-- and right--chiral fermion zero modes tend to a constant and infinity respectively, when far away from the brane. So none of the two zero modes cannot be localized on the brane. If only introducing the second coupling, the effective potentials of the KK modes are modified P\"{o}schl--Teller potentials, which lead to localization of the left-- or right--chiral fermion zero mode as well as to a mass gaps in the mass spectra. When both two couplings are considered, the potentials will have a barrier at each side of the brane, and tend to a positive constant when far away from the brane. Then one of the left-- and right--chiral fermions zero modes can be localized on the brane, and there exists a mass gap in the mass spectra of the fermion KK modes. Then the massive bound and/or resonant KK modes can be localized and/or quasilocalized on the brane.

The organization of this paper is as follows: In Sec. \ref{sec:review}, we first review the Minkowski thick brane generated by two scalar fields. Then, in Sec. \ref{sec:ferLoca}, we investigate localization and quasilocalization of a spin--$1/2$ fermion field on this thick brane by introducing two kinds of scalar--fermion couplings. Finally, our conclusion is given in Sec. \ref{sec:conclusion} together with some discussion on the presented material.

\section{Review of the Minkowski thick braneworld generated by two scalar fields}
\label{sec:review}

In this paper, we consider a Minkowski brane generated by two interacting scalar fields $\phi$ and $\pi$, embedded in a five--dimensional spacetime. The action of such a system is
\begin{eqnarray}
 \label{ActionBrane}
 S=\int d^{5}x\sqrt{-g}\left[ \frac{1}{2\kappa_{5}^{2}}R-\frac{1}{2}(\partial\phi)^{2}
                              -\frac{1}{2}(\partial\pi)^{2}-V(\phi,\pi)
                                 \right],
\end{eqnarray}
where $R$ is the five--dimensional scalar curvature and $\kappa_{5}^{2}=8 \pi G_5$ with $G_5$ the five--dimensional Newton constant. For simplicity, the constant $\kappa_{5}$ is set to $\kappa_5=1$. The line--element of the Minkowski brane is assumed as
\begin{eqnarray}
 \label{LineElementY}
 ds^{2}=\textrm{e}^{2A(y)}\eta_{\mu\nu}dx^{\mu}dx^{\nu}+dy^{2},
\end{eqnarray}
which can also be transformed to the conformally flat one
\begin{eqnarray}
 \label{LineElementZ}
 ds^{2}=\textrm{e}^{2A(z)}(\eta_{\mu\nu}dx^{\mu}dx^{\nu}+dz^{2}),
\end{eqnarray}
by introducing the coordinate transformation $dz=\textrm{e}^{-A(y)}dy$. Here $\textrm{e}^{2A}$ is the warp factor and $\eta_{\mu\nu}$ is the metric of four--dimensional Minkowski spacetime on the brane. $y$ or $z$ denotes the extra dimensional coordinate. The background scalar fields $\phi$ and $\pi$ are assumed to be only the functions of the extra dimensional coordinate.

By considering the metric (\ref{LineElementZ}), the equations of motion generated from action (\ref{ActionBrane}) read as
\begin{eqnarray}\label{EqMotion}
 \frac{1}{2}\phi'^2 + \frac{1}{2}\pi'^2 - \text{e}^{2A} V &=& 6A'^2, \\  \label{EOM1}
 \frac{1}{2}\phi'^2 + \frac{1}{2}\pi'^2 + \text{e}^{2A} V &=& -3A''-3A'^2,\\ \label{EOM2}
 \phi'' + 3A'\phi' &=& \text{e}^{2A}\frac{\partial V}{\partial \phi},\\ \label{EOM3}
 \pi'' + 3A'\pi' &=& \text{e}^{2A}\frac{\partial V}{\partial \pi}, \label{EOM4}
\end{eqnarray}
where the prime denotes the derivative with respect to $z$. The solutions of the thick brane can be found by following the superpotential method \cite{DeWolfe}. By supposing $V(\phi,\pi)=\frac{1}{2}\text{e}^{-\frac{2\pi}{\sqrt{3}}}\left[(\frac{\partial W}{\partial\phi})^{2}-W^{2}\right]$ and considering a specific superpotential function $W(\phi)=v a\phi(1-\frac{\phi^{2}}{3 v^{2}})$ \cite{two-scalarbrane1,KRthickbrane}, one solution is obtained
\begin{eqnarray}
  A(z)&=&-\frac{v^2}{9}\bigg[\ln(\cosh^2(az))
                        +\frac{1}{2}\tanh^2(az)\bigg], \label{warpfactor}\\
  \phi(z)&=&v\tanh(az),\label{kink}\\
  \pi(z)&=&\sqrt{3} \;A(z), \label{dilaton} \\
  V(\phi,\pi)&=&\frac{a^2}{18 v^2} \textrm{e}^{-\frac{2\pi}{\sqrt{3}}}
                \big[9v^4-9v^{2}(v^2+2)\phi^{2}+3(2v^2+3)\phi^{4}-\phi^{6}\big],
\end{eqnarray}
where $v$ and $a$ are positive constants. It can be seen that the solution for $\phi$ is a kink and $\pi$ is the dilaton field, which are odd and even functions of the extra dimensional coordinate $z$, respectively. The energy density of the thick brane is written as follows
\begin{eqnarray}\label{T00}
 T_{00}(z)&=&\frac{1}{2}(\phi'^{2}+\pi'^{2})+\textrm{e}^{2A}V(\phi,\pi)\nonumber\\
          &=&\frac{a^2 v^2}{27} \Big[v^{2}\textrm{sech}^{6}(a z)
              +3 (v^{2}+9)\textrm{sech}^{4}(a z)-4v^2\Big].
\end{eqnarray}
From the above expression (\ref{T00}), we can see that the energy density has a maximum value $T_{00}^{\textrm{max}}=a^{2}v^{2}$ at $z=0$ and tends to the minimum value $T_{00}^{\textrm{min}}=-\frac{4}{27}a^{2}v^{2}$ when $z\rightarrow \pm\infty$, so the Minkowski brane locates at $z=0$. Localization of gravity, scalar, vector, and Kalb-Ramond fields was investigated in previous papers \cite{twofieldbrane,deformedBrane}. So in the next section, localization of a spin-$1/2$ fermion with two types of couplings between  the fermion and background scalar fields will be discussed.

\section{\label{sec:ferLoca} Localization of a bulk fermion on the brane}

In this section, we will investigate localization of a bulk spin-$1/2$ fermion on the brane generated by the scalar fields. In five--dimensional spacetime, fermions are four--component spinors and their Dirac structure can be described by $\Gamma^M= e^M_{~\bar{M}} \Gamma^{\bar{M}}$ with $e^M_{~\bar{M}}$ being the vielbein and $\{\Gamma^M,\Gamma^N\}=2g^{MN}$. In this section, $\bar{M}, \bar{N}, \cdots =0,1,2,3,5$ and $ M, N, \cdots =0,1,2,3,5$ denote the five--dimensional local Lorentz indices and spacetime coordinates, respectively, and $\Gamma^{\bar{M}}$ are the gamma matrices in five--dimensional flat spacetime. Our action describing a massless Dirac fermion coupled with the background scalars $\phi$ and $\pi$ in five--dimensional space--time is assumed as follows
\begin{eqnarray}
 S_{\frac{1}{2}} = \int d^5 x \sqrt{-g} \left[\bar{\Psi} \Gamma^M
          \left(\partial_M+\omega_M\right) \Psi
          -\eta\bar{\Psi}f(\phi,\pi)\Psi
          +\lambda\bar{\Psi}\Gamma^{M}\partial_{M} g(\phi,\pi)\gamma^{5}\Psi\right]. \label{DiracAction}
\end{eqnarray}
Here $\omega_M$ is the spin connection defined as $\omega_M=\frac{1}{4} \omega_M^{\bar{M} \bar{N}} \Gamma_{\bar{M}}\Gamma_{\bar{N}}$. Both the usual Yukawa coupling $-\eta\bar{\Psi}f(\phi,\pi)\Psi$ and a new kind of coupling between the
Dirac fermion and background scalar fields \cite{liuPRD89} $\lambda\bar{\Psi}\Gamma^{M}\partial_{M} g(\phi,\pi)\gamma^{5}\Psi$ are introduced. In this paper, we assume that both coupling constants $\eta$ and $\lambda$ are positive.

Considering the conformally flat metric (\ref{LineElementZ}), the non--vanishing components of the spin connection are $\omega_\mu =\frac{1}{2}A'(z) \gamma_\mu \gamma_5$.
Thus, the equation of motion corresponding to the action (\ref{DiracAction})
can be written as
\begin{eqnarray}
 \left[ \gamma^{\mu}\partial_{\mu}
         + \gamma^5 \left(\partial_z  +2  A'(z) \right)
         -\eta\textrm{e}^{A}f(\phi,\pi)+\lambda g'(\phi,\pi)
 \right] \Psi =0. \label{DiracEq1}
\end{eqnarray}

In order to investigate the above five--dimensional Dirac equation (\ref{DiracEq1}) and  write the spinor in terms of four--dimensional effective fields, we make the general chiral decomposition:
\begin{equation}
 \Psi= \text{e}^{-2A}\Big(\sum_n\psi_{Ln}(x) L_n(z)
 +\sum_n\psi_{Rn}(x) R_n(z)\Big),
\end{equation}
where $\psi_{Ln}(x)=-\gamma^5 \psi_{Ln}(x)$ and $\psi_{Rn}(x)=\gamma^5 \psi_{Rn}(x)$ are the left-- and right--chiral components of a four--dimensional Dirac field,
respectively. Hence, we assume that $\psi_{Ln}(x)$ and
$\psi_{Rn}(x)$ satisfy the four--dimensional Dirac equations
\begin{subequations}\label{CoupleEq0}
\begin{eqnarray}
 \label{DiracEq4D}
 \gamma^{\mu}\partial_{\mu}\psi_{Ln}(x)&=& m_{n}\psi_{Rn}(x),  \\
 \gamma^{\mu}\partial_{\mu}\psi_{Rn}(x)&=& m_{n}\psi_{Ln}(x).
\end{eqnarray}
\end{subequations}
Then the KK modes $L_{n}(z)$ and $R_{n}(z)$ should satisfy the following
coupled equations:
\begin{subequations}\label{CoupleEq1}
\begin{eqnarray}
 \left[\partial_{z}
                  + \eta\textrm{e}^A f(\phi,\pi) -\lambda g'(\phi,\pi) \right]L_n(z)   &=&  ~~m_n R_n(z), \label{CoupleEq1a}  \\
 \left[\partial_z
                 - \eta\textrm{e}^A f(\phi,\pi) +\lambda g'(\phi,\pi) \right]R_n(z)
                  &=&  - m_n L_n(z). \label{CoupleEq1b}
\end{eqnarray}
\end{subequations}
The above coupled equations can be recast into
\begin{subequations}\label{emieq}
\begin{eqnarray}
 Q^{\dag}Q L_{n}(z)&=&m_{n}^{2}L_{n}(z), \\
 QQ^{\dag} R_{n}(z)&=&m_{n}^{2}R_{n}(z),
\end{eqnarray}
\end{subequations}
where the operator $Q$ is defined as $Q=\partial_{z}+\eta\textrm{e}^{A}f(\phi,\pi)-\lambda g'(\phi,\pi)$. According to the supersymmetric quantum mechanics, there are no tachyon fermion KK modes with negative mass square $m_{n}^{2}$. The above equations (\ref{emieq}) can also be rewritten as the following Schr\"{o}dinger--like equations for the left-- and right--chiral KK modes of fermions:
\begin{subequations} \label{SchEqFermion}
\begin{eqnarray}
  \big(-\partial^2_z + V_L(z) \big)L_n
            &=&m_{L_n}^{2} L_n,~~
   \label{SchEqLeftFermion}  \\
  \big(-\partial^2_z + V_R(z) \big)R_n
            &=&m_{R_n}^{2} R_n,
   \label{SchEqRightFermion}
\end{eqnarray}
\end{subequations}
where the effective potentials are given by
\begin{subequations}\label{Vfermion}
\begin{eqnarray}
  V_L(z)&=& \eta^{2}\textrm{e}^{2A}f^{2}(\phi,\pi)+\lambda^{2}g'^{~2}(\phi,\pi)
           -\eta\textrm{e}^{A}[A'f(\phi,\pi)+f'(\phi,\pi)]\nonumber\\
         &&~~
         -2\eta\lambda\textrm{A}f(\phi,\pi)g'(\phi,\pi)+\lambda g''(\phi,\pi), \label{VL}\\
  V_R(z)&=&  \eta^{2}\textrm{e}^{2A}f^{2}(\phi,\pi)+\lambda^{2}g'^{~2}(\phi,\pi)
           +\eta\textrm{e}^{A}[A'f(\phi,\pi)+f'(\phi,\pi)]\nonumber\\
         &&~~
         -2\eta\lambda\textrm{A}f(\phi,\pi)g'(\phi,\pi)-\lambda g''(\phi,\pi). \label{VR}
\end{eqnarray}
\end{subequations}
For the purpose of getting the standard four--dimensional action for
a massless chiral fermion and a series of massive fermions
\begin{eqnarray}
 S_{\frac{1}{2}} &=& \int d^5 x \sqrt{-g} ~\bar{\Psi}
     \left[ \Gamma^M (\partial_M+\omega_M)
     -\eta f(\phi,\pi)+\lambda\Gamma^{M}\partial_{M} g(\phi,\pi)\gamma^{5}\right]\Psi
      \nonumber \\
  &=&\sum_{n}\int d^4 x
    ~\bar{\psi}_{n}
      \left[\gamma^{\mu}\partial_{\mu}
        -m_{n}\right]\psi_{n},~~~
\end{eqnarray}
the following orthonormalization conditions for $L_{n}$ and $R_{n}$
are needed:
\begin{eqnarray}
 \int_{-\infty}^{+\infty} L_m L_ndz
   &=& \delta_{mn}, \label{orthonormalityFermionL} \\
 \int_{-\infty}^{+\infty} R_m R_ndz
   &=& \delta_{mn}, \label{orthonormalityFermionR}\\
 \int_{-\infty}^{+\infty} L_m R_ndz
   &=& 0. \label{orthonormalityFermionLR}
\end{eqnarray}

By setting $m_n=0$ in Eqs. (\ref{CoupleEq1}), we can obtain the left-- and right--chiral KK fermion zero modes
\begin{subequations}
\begin{eqnarray}
  L_0&\propto & \textrm{e}^{\lambda g(\phi,\pi)-\eta \int e^{A} f(\phi,\pi)dz}, \label{zerol}\\
  R_0&\propto & \textrm{e}^{-\lambda g(\phi,\pi)+\eta \int e^{A} f(\phi,\pi)dz}. \label{zeror}
\end{eqnarray}
\end{subequations}
It is impossible to make both massless left-- and right--chiral KK fermion modes to be localized on the brane at the same time, since when one is normalizable, the other one is not.

From Eqs. (\ref{SchEqFermion}) and (\ref{Vfermion}), it is clear that, if we do not introduce the coupling term in the action (\ref{DiracAction}), i.e., $\eta=0$ and $\lambda=0$, the effective potentials for left-- and right--chiral KK modes $V_{L,R}(z)$ will vanish and both left-- and right--chiral fermions cannot be localized on the thick brane. Moreover, if we demand $V_{L}(z)$ and $V_{R}(z)$ to be $Z_{2}$-even with respect to the extra dimension $z$, then the coupling functions $f(\phi, \pi)$ and $g(\phi,\pi)$ must be odd and even functions of $z$, respectively. The simplest choice is that $f(\phi,\pi)=\phi$ and $g(\phi,\pi)=\pi$, since the kink scalar field $\phi$ and dilaton field $\pi$ are odd and even function of $z$. In this paper, we want to compare the effect of localization of a bulk fermion with two types of couplings.

\begin{figure*}[htb]
\begin{center}
\subfigure[$V_{L}(z), \lambda=5$]{\label{fig_Vldfeta}
\includegraphics[width=7cm]{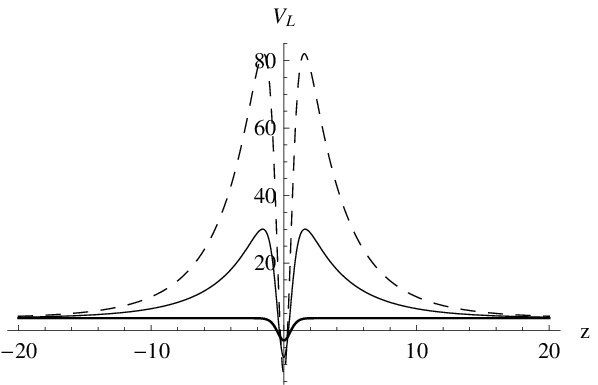}}
\subfigure[$V_{R}(z), \lambda=5$]{\label{fig_Vrdfeta}
\includegraphics[width=7cm]{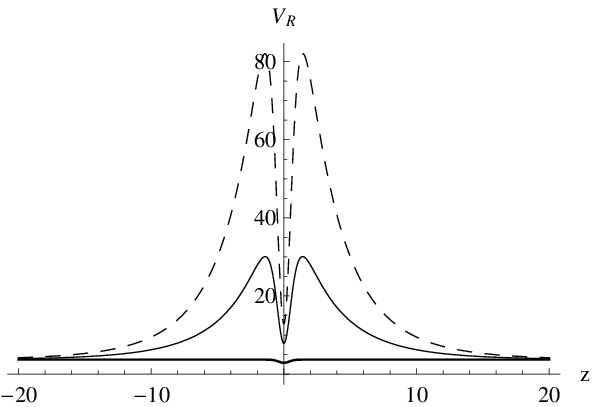}}
\end{center}\vskip -5mm
\caption{The shapes of the effective potentials of the left-- and right--chiral fermions $V_{L,R}(z)$ for different values of the parameter $\eta$.
The parameter $\eta$ is set to $\eta=0$ for the thick line, $\eta=5$ for the thin line, and $\eta=10$ for the dashed line. Other parameters are set to $v=1, a=1$ and $\lambda=5$.}
 \label{fig_Vlrdifeta}
\end{figure*}

\begin{figure*}[htb]
\begin{center}
\subfigure[$V_{L}(z), \eta=5$]{\label{fig_Vldflambda}
\includegraphics[width=7cm]{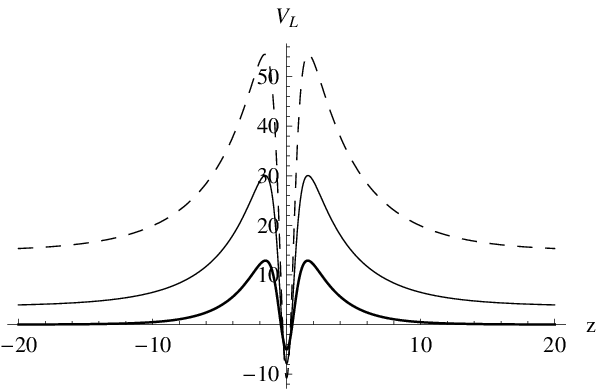}}
\subfigure[$V_{R}(z), \eta=5$]{\label{fig_Vrdflambda}
\includegraphics[width=7cm]{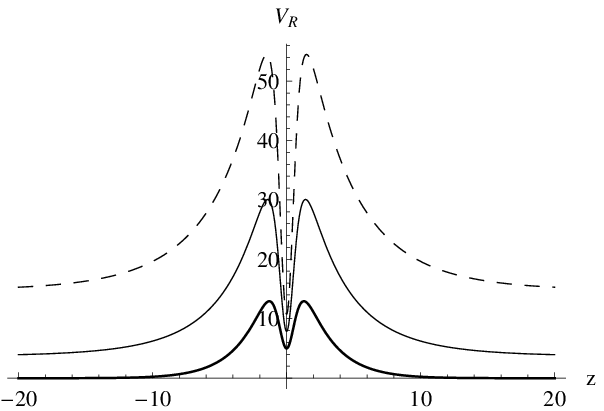}}
\end{center}\vskip -5mm
\caption{The shapes of the effective potentials of the left-- and right--chiral fermions $V_{L,R}(z)$ for different values of the parameter $\lambda$.
The parameter $\lambda$ is set to $\lambda=0$ for the thick line, $\lambda=5$ for the thin line, and $\lambda=10$ for the dashed line. Other parameters are set to $v=1, a=1$ and $\eta=5$.}
 \label{fig_Vlrdiflambda}
\end{figure*}

By considering the solution of the Minkowski brane (\ref{warpfactor}), (\ref{kink}) and (\ref{dilaton}), the effective potentials (\ref{Vfermion}) can be expressed as follows
\begin{eqnarray}
 V_{L}(z)&=& \frac{1}{27}v\textrm{e}^{-\frac{1}{9}v^{2}\tanh^{2}az}
               (\cosh az)^{-\frac{4v^{2}}{9}}
              \nonumber\\
          &&~  \times\Big[3a\eta\textrm{e}^{\frac{1}{18}v^{2}\tanh^{2}az}
               (\cosh az)^{2(\frac{v^{2}}{9}-1)} (v^{2}(2\sqrt{3}\lambda+1)(2+\cosh2az)\tanh^{2}az-9)
              \nonumber\\
          &&~~   +a^{2}v\lambda\textrm{e}^{\frac{1}{9}v^{2}\tanh^{2}az}
                (\cosh az)^{\frac{4v^{2}}{9}}
                \textrm{sech}^{4}az(v^{2}\lambda(\sinh2az+\tanh az)^{2}-9\sqrt{3})
                \nonumber\\
           &&~~       +27v \eta^{2} \tanh^{2}az \Big], \label{VL}
\end{eqnarray}
\begin{eqnarray}
 V_{R}(z)&=& \frac{1}{27}v\textrm{e}^{-\frac{1}{9}v^{2}\tanh^{2}az}
               (\cosh az)^{-\frac{4v^{2}}{9}}
              \nonumber\\
          &&~  \times\Big[3a\eta\textrm{e}^{\frac{1}{18}v^{2}\tanh^{2}az}
               (\cosh az)^{2(\frac{v^{2}}{9}-1)} (v^{2}(2\sqrt{3}\lambda-1)(2+\cosh2az)\tanh^{2}az+9)
              \nonumber\\
          &&~~   +a^{2}v\lambda\textrm{e}^{\frac{1}{9}v^{2}\tanh^{2}az}
                (\cosh az)^{\frac{4v^{2}}{9}}
                \textrm{sech}^{4}az(v^{2}\lambda(\sinh2az+\tanh az)^{2}+9\sqrt{3})
                \nonumber\\
           &&~~       +27v \eta^{2} \tanh^{2}az \Big]. \label{VR}
\end{eqnarray}
Here we need to analyze the behavior of the potentials near $z=0$ and as $z\rightarrow\pm\infty$, respectively. From the above expressions (\ref{VL}) and (\ref{VR}), it is easy to obtain
\begin{eqnarray}
 V_L (z=0)&=&-\frac{1}{3}a v(3\eta+\sqrt{3} av\lambda),\\
 V_R (z=0)&=&~~\frac{1}{3}a v(3\eta+\sqrt{3} av\lambda),
\end{eqnarray}
and
\begin{eqnarray}
 V_{L,R}~'(z=0)&=&0,\\
 V_{L}~''(z=0)&=& \frac{1}{3}a^{2}v\Big[ 2a^{2}v^{3}\lambda^{2}
                  +4\sqrt{3}av(a+v\eta)\lambda+6v\eta^{2}+3a(2+v^{2})\eta\Big],\\
 V_{R}~''(z=0)&=& \frac{1}{3}a^{2}v\Big[ 2a^{2}v^{3}\lambda^{2}
                  -4\sqrt{3}av(a+v\eta)\lambda+6v\eta^{2}-3a(2+v^{2})\eta\Big].
\end{eqnarray}
Since the parameters $a$, $v$, $\eta$, and $\lambda$ are all positive, $V_{L}''(z=0)>0$ and $V_{L}(z)$ has a local minimum value at $z=0$.
As $z\rightarrow\pm\infty$, the both potentials tend to a constant
\begin{eqnarray}
 V_{L,R}(z\rightarrow\pm\infty)=\frac{4}{27}a^{2}v^{4}\lambda^{2}.
\end{eqnarray}
The shapes of the effective potentials for the left-- and right--chiral fermions are plotted in  Figs.~\ref{fig_Vlrdifeta} and \ref{fig_Vlrdiflambda}.

\begin{figure}[htb]
\begin{center}
\includegraphics[width=7cm]{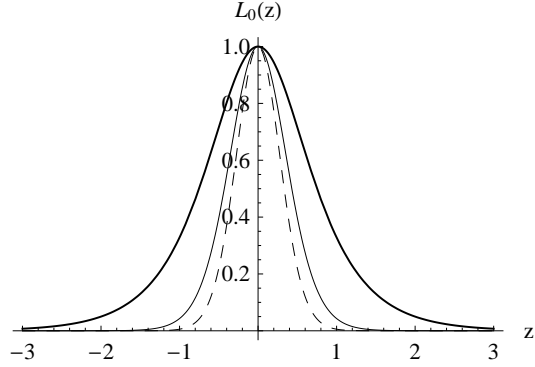}
\end{center}\vskip -5mm
\caption{The shapes of the zero mode of the left--chiral fermion $L_{0}(z)$ for different values of the parameter $\eta$. The coupling coefficient $\eta$ is set to $\eta=0$, $\eta=5$, and $\eta=10$ for thick, thin, and dashing lines, respectively. Other parameters are set to $v=1$, $a=1$, $C_{L}=1$ and $\lambda=5$.}
 \label{fig_L0eta}
\end{figure}

\begin{figure}[htb]
\begin{center}
\subfigure[$-3\leq z\leq3$]{\label{fig_L0lambda_a}
\includegraphics[width=7cm]{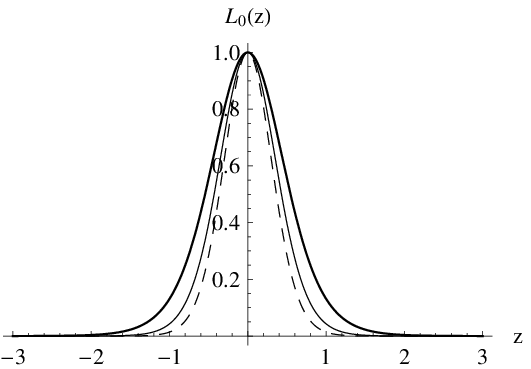}}
\subfigure[$10\leq z\leq30$]{\label{fig_L0lambda_b}
\includegraphics[width=7cm]{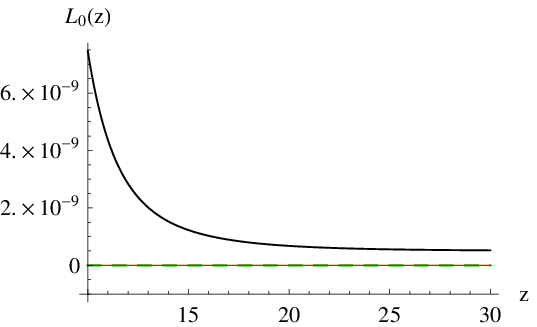}}
\end{center}\vskip -5mm
\caption{The shapes of the zero mode of the left--chiral fermion $L_{0}(z)$ for different values of the parameter $\lambda$. The coupling coefficient $\lambda$ is set to $\lambda=0$, $\lambda=5$, and $\lambda=10$ for thick, thin, and dashing lines, respectively. Other parameters are set to $v=1$, $a=1$, $C_{L}=1$, and $\eta=5$.}
 \label{fig_L0lambda}
\end{figure}

It is clear that, with the positive coupling coefficients $\eta$ and $\lambda$, only the potential for left--chiral fermion KK modes has negative part around $z=0$. So only the left--chiral fermion zero mode may be localized on the brane. The expression of the zero mode can be written as
\begin{eqnarray}
 L_{0}(z)&=& C_{L}\textrm{e}^{\lambda\pi
                  -\eta\int_{0}^{z}\textrm{e}^{A(\bar{z})}\phi(\bar{z})d\bar{z}}
                 \nonumber\\
          &=& C_{L} (\cosh az)^{-\frac{2\sqrt{3}}{9}v^{2}\lambda}
               \textrm{e}^{-\frac{\sqrt{3}}{18}v^{2}\lambda\tanh^{2}az
                -v\eta\int_{0}^{z} (\cosh a\bar{z})^{-\frac{2v^{2}}{9}}
               (\tanh a\bar{z})\textrm{e}^{-\frac{1}{18}v^{2}\tanh^{2}a\bar{z}}d\bar{z}},
\end{eqnarray}
where $C_L$ is a constant. Here the explicit analytic expression of the left--chiral fermion zero mode cannot be obtained. However, from the asymptotic behavior of the warp factor $A(z)$, kink scalar field $\phi(z)$, and dilaton scalar field $\pi(z)$ as $z\rightarrow\pm\infty$, the asymptotic behavior of the zero mode can be analyzed:
\begin{eqnarray}
 L_{0}(z\rightarrow\pm\infty)\rightarrow C_{L}2^{\frac{2\sqrt{3}}{9}v^{2}\lambda}
                             \exp(-\frac{9\eta}{2av}2^{\frac{2v^{2}}{9}}
                                  \textrm{e}^{-\frac{v^{2}}{18}}
                                  -\frac{\sqrt{3}}{18}v^{2}\lambda
                                  -\frac{2\sqrt{3}}{9}v^{2}\lambda az).
\end{eqnarray}
Then we can conclude from the above expression that, the zero mode of left--chiral fermion tends to a constant $C_{L}\textrm{e}^{-\frac{9\eta}{2av}2^{\frac{2v^{2}}{9}}\textrm{e}^{-\frac{v^{2}}{18}}}$ when the coupling coefficient $\lambda=0$, and it cannot be localized on the brane. For the purpose of localizing the zero mode of left--chiral fermion, the coupling term
$\lambda\bar{\Psi}\Gamma^{M}\partial_{M}\pi\gamma^{5}\Psi$ in the action (\ref{DiracAction}) must be introduced. We also show the shapes of the left--chiral fermion zero mode for for different values of the parameters $\eta$ and $\lambda$ in Figs. \ref{fig_L0eta} and \ref{fig_L0lambda}, respectively. In Fig. \ref{fig_L0lambda_b}, it is clear that, when $\lambda=0$, the left--chiral fermion zero mode tends to a constant as $z\rightarrow\pm\infty$, and it cannot be localized on the brane. Thus, we set the coupling coefficient $\lambda>0$. So $V_{L}(z\rightarrow\pm\infty)=\frac{4}{27}a^{2}v^{4}\lambda^{2}>0$, and there is a massless bound state (the left--chiral fermion zero mode) at least, and continuous spectra of the left-- and right--chiral fermion KK modes staring from $m^2 >\frac{4}{27}a^{2}v^{4}\lambda^{2}$, which indicates the presence of a mass gap in the fermion KK modes spectra.

Next, the effect of the two types of couplings between the fermion and scalar fields for localization of the fermion KK modes will be investigated. From Fig. \ref{fig_Vlrdifeta}, we can see that, when $\lambda>0$ and $\eta=0$, the potentials $V_{L,R}(z)$ become the P\"{o}schl--Teller potentials, and the bound states may exist, i.e., the KK modes can be localized on the brane. When $\lambda>0$ and $\eta>0$, there are two symmetrical potential barriers at both sides of the origin of the extra dimension, and they increase with the coupling coefficient $\eta$. Generally, this type potential implies that there may exist bound states when $0{\leq}m^{2}{<}\frac{4}{27}a^{2}v^{4}\lambda^{2}$, and may also exist resonant states when $\frac{4}{27}a^{2}v^{4}\lambda^{2}{<}m^{2}{\leq}V_{L,R}^{\textrm{max}}$ ($V_{L,R}^{\textrm{max}}$ is the maximum value of $V_{L,R}$), which indicates that the KK modes are quasilocalized on the brane. The bound states tend to zero when far away from the brane along the extra dimension, and they can be normalized. However, the resonant states tend to plane waves when far away from the brane, and cannot be normalized. As in Ref. \cite{Liu0907.0910}, the relative probability function of a resonance on the thick brane is defined as follows
\begin{equation}
 P_{L,R}(m^{2})=\frac{\int_{-z_b}^{z_b} |L(z),R(z)|^2 dz}
                 {\int_{-z_{max}}^{z_{max}} |L(z),R(z)|^2 dz},
\label{ProbabilityLR}
\end{equation}
where $2z_{b}$ is about the width of the thick brane, and $z_{max}$ is set to $z_{max}=10z_{b}$. It is clear that when $m^2 \gg V_{L,R}^{\textrm{max}}$, the fermion KK modes are approximately taken as plane waves and the value of $P_{L,R}(m^2)$ will tend to $1/10$. As in Ref. \cite{PRL84}, the lifetime $\tau$ is estimated to $\tau\sim\Gamma^{-1}$, with $\Gamma=\delta m$ being the full width at half maximum of the peak.

\begin{table*}[h]
\begin{center}
\renewcommand\arraystretch{1.3}
\begin{tabular}
 {|l|c|c|c|c|c|c|c|c|c|c|}
 \hline
 $\lambda$ &$\eta$ & Chiral & Height of $V_{\text{L,R}}$ & $n$  &
  Bound or resonant & $m^{2}$  &  $m$  &  $\Gamma$  & $\tau$    \\
    \hline\hline
 5 & $0$  & Left &  $V_{L}^{\textrm{PT}}(\infty)=3.7037$
     &  0   & bound state & 0 & 0 & 0 & $\infty$ \\
    \cline{3-10}
  &   &  Right & $V_{R}^{\textrm{PT}}(\infty)=3.7037$
     &  ---   &  ---  &  ---  &  ---  &  ---  &  ---  \\
   \cline{2-10}
  & $5$  &  Left & $V_{L}^{\text{max}}=30.048$
     &  $0$   & bound state  & 0 &  0  &  0  &  $\infty$  \\
    \cline{5-10}
  &  &   &
    &  $1$   & resonant state  & 14.039 &  3.75  &  $1.30\times 10^{-6}$  & $7.67\times10^{5}$  \\
    \cline{5-10}
  &  &  &
     &  $2$  & resonant state  &  24.405  &  4.94    & $6.17\times10^{-3}$   &  158.67        \\
     \cline{5-10}
 &  &  &
     &  $3$  & resonant state  &  30.380  &  5.51    &  0.163    &  6.119        \\
     \cline{3-10}
 &  &  Right & $V_{R}^{\text{max}}=30.071$
     &  $1$   & resonant state   & 14.039 &  3.75  &  $1.33\times 10^{-6}$  & $7.52\times10^{5}$  \\
    \cline{5-10}
  &  &  &
     &  $2$  & resonant state  &  24.405  &  4.94    & $6.16\times10^{-3}$   &  162.321        \\
     \cline{5-10}
  & &  &
     &  $3$  & resonant state  &  30.329  &  5.51    &  0.168    &  5.939        \\
   \cline{2-10}
 & $10$  &  Left & $V_{L}^{\text{max}}=82.009$
     &  $0$   & bound state  & 0 &  0  &  0  &  $\infty$  \\
    \cline{5-10}
 &  &   &
    &  $1$   & resonant state  & 24.134 &  4.91  &  $6.93\times 10^{-13}$  & $1.45\times10^{12}$  \\
    \cline{5-10}
 &  &  &
     &  $2$  & resonant state  &  44.911  &  6.70    & $2.50\times10^{-7}$   & $3.999\times10^{6}$        \\
     \cline{5-10}
 &  &  &
     &  $3$  & resonant state  &  62.117  &  7.88    &  $2.84\times10^{-4}$    & $3.52\times10^{3}$       \\
     \cline{5-10}
 &  &  &
     &  $4$  & resonant state  &  75.233  &  8.67   &  $1.83\times10^{-2}$
     & $54.588$       \\
     \cline{5-10}
 &  &  &
     &  $5$  & resonant state  &  83.240  &  9.12    &  0.213    &  4.690   \\
     \cline{3-10}
 &  &  Right & $V_{R}^{\text{max}}=82.023$
     &  $1$   & resonant state   & 24.134 &  4.91  &  $6.98\times 10^{-13}$  & $1.43\times10^{12}$  \\
    \cline{5-10}
 &  &  &
     &  $2$  & resonant state  &  44.911  &  6.70    & $2.48\times10^{-7}$   & $4.03\times10^{6}$       \\
     \cline{5-10}
 &  &  &
     &  $3$  & resonant state  &  62.117  &  7.88    & $2.85\times10^{-4}$   &
       $3.51\times10^{3}$        \\
   \cline{5-10}
 &  &  &
     &  $4$  & resonant state  &  75.231  &  8.67    & $1.84\times10^{-2}$   &
       54.212        \\
   \cline{5-10}
 &  &  &
     &  $5$  & resonant state  &  83.185  &  9.12    & 0.232   &
       4.311        \\
   \hline
\end{tabular}
\end{center}
\caption{The mass, width, and lifetime of bound or resonant KK modes of the fermions.
The parameters are set to $a=1$, $v=1$, $\lambda=5$, and $\eta=0, 5, 10$. Here $n$ is the level of KK modes with the corresponding $m^2$ from small to large.} \label{tableLR1}
\end{table*}

\begin{figure*}[htb]
\begin{center}
\subfigure[ Life--chiral]{\label{fig_PL55}
\includegraphics[width=7cm]{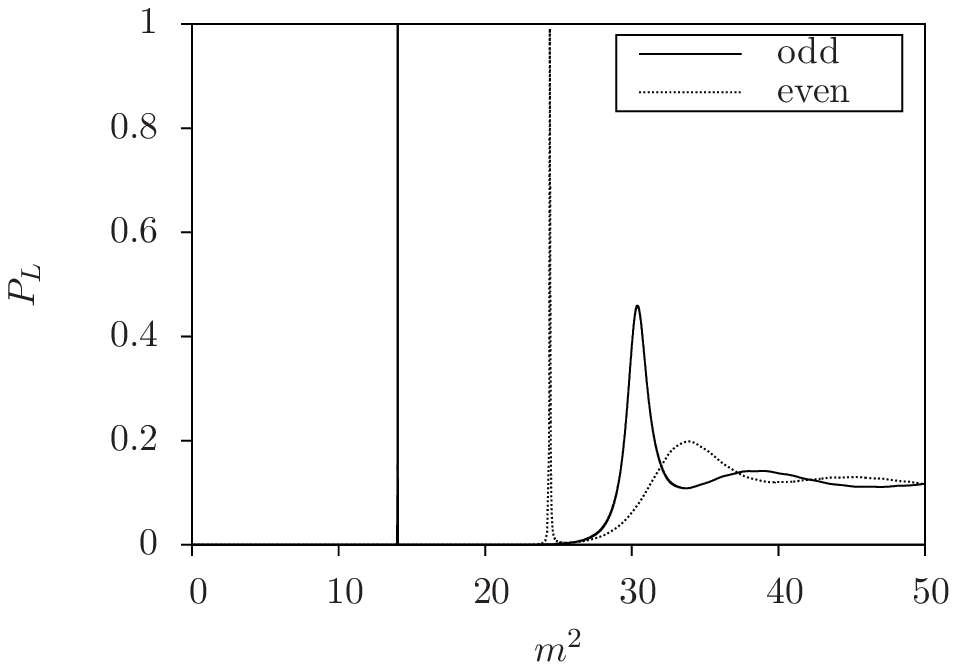}}
\subfigure[ Right--chiral]{\label{fig_PR55}
\includegraphics[width=7cm]{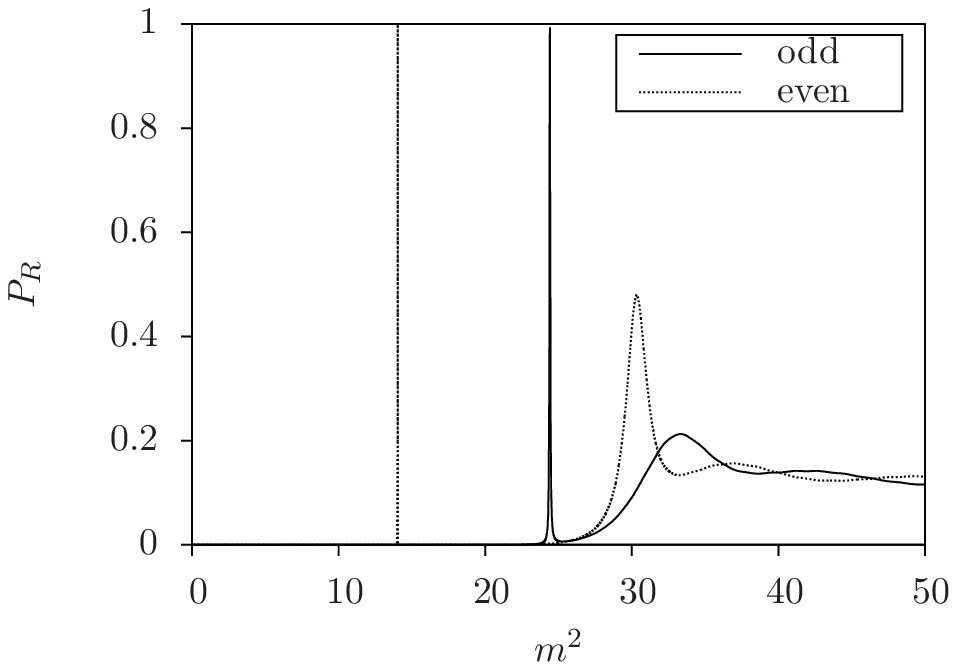}}
\end{center}\vskip -5mm
\caption{The shapes of the relative probability $P_{L,R}(m^{2})$ for the left-- and right--chiral fermion resonant KK modes, with the parameters $a=1$, $v=1$, $\lambda=5$, and $\eta=5$.}
 \label{fig_PLR55}
\end{figure*}

\begin{figure}[htb]
\begin{center}
\subfigure[$n=0$, bound state]{\label{fig_L055}
\includegraphics[width=7cm]{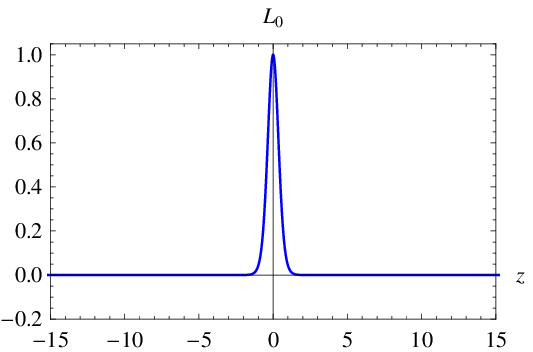}}
\subfigure[$n=1$, resonant state]{\label{fig_L155}
\includegraphics[width=7cm]{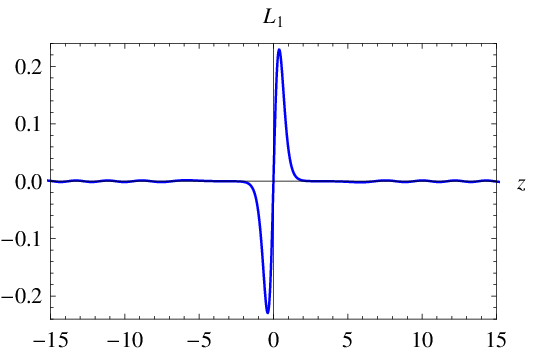}}
\subfigure[$n=2$, resonant state]{\label{fig_L255}
\includegraphics[width=7cm]{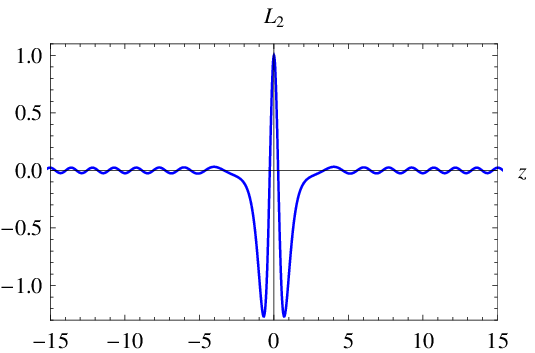}}
\subfigure[$n=3$, resonant state]{\label{fig_L355}
\includegraphics[width=7cm]{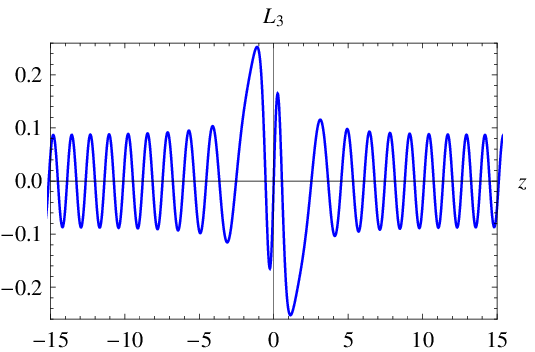}}
\end{center}\vskip -5mm
\caption{The shapes of the wave functions for the left--chiral fermion KK modes $L_n$, with the parameters $a=1$, $v=1$, $\lambda=5$, and $\eta=5$.}
 \label{fig_L55}
\end{figure}

\begin{figure}[htb]
\begin{center}
\subfigure[$n=1$, resonant state]{\label{fig_R155}
\includegraphics[width=7cm]{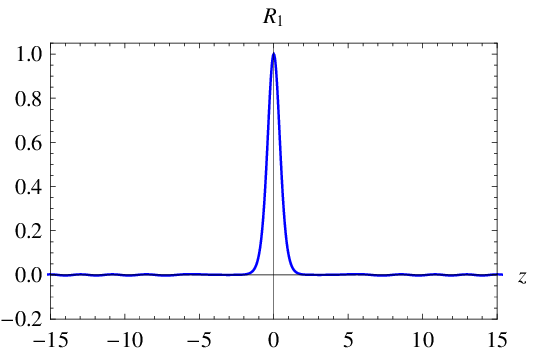}}
\subfigure[$n=2$, resonant state]{\label{fig_R255}
\includegraphics[width=7cm]{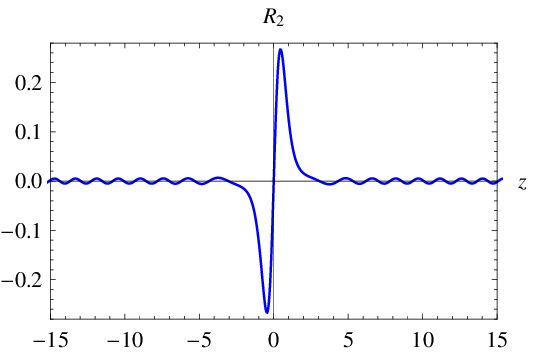}}
\subfigure[$n=3$, resonant state]{\label{fig_R355}
\includegraphics[width=7cm]{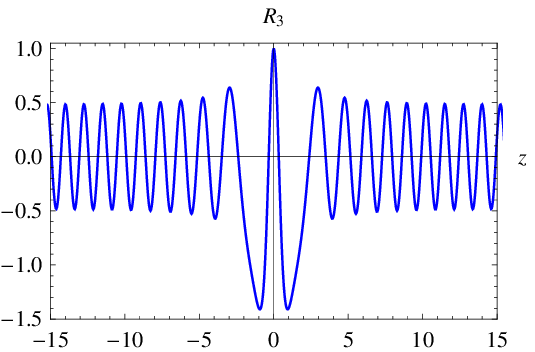}}
\end{center}\vskip -5mm
\caption{The shapes of the wave functions for the right--chiral fermion KK modes $R_n$, with the parameters $a=1$, $v=1$, $\lambda=5$, and $\eta=5$.}
 \label{fig_R55}
\end{figure}

Equations (\ref{SchEqFermion}) can be solved by numerical method, and we will set the coupling coefficients $\lambda$ and $\eta$ as different values respectively for comparing two types of couplings. When $\lambda=5$ and $\eta=0$, only the left--chiral fermion zero mode (bound state) can be localized on the brane. However, for the right--chiral fermion KK modes, there is no bound state or resonant state. When $\lambda=5$ and $\eta=5$, for the left--chiral fermion KK modes, there is only one bound zero mode and three resonant KK modes; but for right--chiral fermion KK modes, there are only three resonant KK modes. The profiles of the relative probability $P_{L,R}(m^{2})$ corresponding to coupling coefficients $\lambda=5$ and $\eta=5$ are shown in Fig. \ref{fig_PLR55}. In these figures, each peak corresponds to a resonant state, and the left-- and right--chiral fermion KK modes are shown in Figs. \ref{fig_L55} and \ref{fig_R55}. For the left--chiral fermion KK modes, it can be seen that the first massive KK mode is an odd--parity wave function, and the second one
has even parity. However, for the right--chiral fermion KK modes, the first massive KK mode has even parity and the second one has odd parity.  This is held for any $n$-th fermion KK mode, namely, the parities of the $n$--th left-- and right--chiral KK modes are opposite. In fact, this conclusion is originated from the coupled equations of the left-- and right--chiral fermions. For the case of $\lambda=5$ and $\eta=10$, the conclusion is similar to that of $\lambda=5$, $\eta=5$. The mass, width, and lifetime of the left-- and right--chiral fermion KK modes with different values of $\eta$ are listed in Table \ref{tableLR1}. It can be seen that the mass and lifetime of the left-- and right--chiral fermion resonances are almost the same, thus the formation of the four-dimensional massive Dirac fermions can be realized \cite{0901.3543}. So we can summarize that the four--dimensional massless left--chiral fermion can be localized on the brane, and the four-dimensional massive Dirac fermions can also be obtained, which are consisted of the pairs of coupled left-- and right--chiral KK modes with different parities. On the other hand, the total number of resonant KK modes increases with the coupling coefficient $\eta$.

\begin{table*}[h]
\begin{center}
\renewcommand\arraystretch{1.3}
\begin{tabular}
 {|l|c|c|c|c|c|c|c|c|c|}
 \hline
 $\lambda$  & $\eta$ & Height of $V_{\text{L}}$ & $n$  &
  Bound or resonant & $m^{2}$  &  $m$  &  $\Gamma$  & $\tau$    \\
    \hline\hline
 $10$  & $0$  & $V_{L}^{\textrm{PT}}(\infty)=14.8148$
        &  0   & bound state & 0 & 0 & 0 & $\infty$ \\
    \cline{4-9}
    &   &
        &  1   & bound state & 9.4517 & 3.0744 & 0 & $\infty$  \\
     \cline{2-9}
     & $5$   &  $V_{L}^{\text{max}}=54.488536$
        & 0    & bound state & 0 & 0 & 0 & $\infty$ \\
    \cline{4-9}
    &   &
        &  1  & resonant state & 19.7422 & 4.443 & $3.00\times10^{-15}$
              & $3.75\times10^{14}$  \\
    \cline{4-9}
    &   &
        &  2  & resonant state & 35.7677 & 5.981 & $3.41\times10^{-6}$
               & $2.94\times10^{5}$  \\
    \cline{4-9}
    &   &
        &  3  & resonant state & 47.7286 & 6.909 & $5.38\times10^{-3}$
               & $1.86\times10^{2}$  \\
    \cline{4-9}
    &   &
        &  4  & resonant state & 54.7649 & 7.400 & $0.135$
               & $7.390$  \\

   \cline{2-9}
     & $10$   &  $V_{L}^{\text{max}}=119.99$
        & 0    & bound state & 0 & 0 & 0 & $\infty$ \\
    \cline{4-9}
    &   &
        &  1  & resonant state & 29.8414 & 5.4627 & $3.56\times10^{-15}$
              & $2.81\times10^{14}$  \\
    \cline{4-9}
    &   &
        &  2  & resonant state & 56.2141 & 7.4976 & $1.04\times10^{-11}$
              & $9.59\times10^{10}$  \\
    \cline{4-9}
    &   &
        &  3  & resonant state & 78.9737 & 8.8867 & $4.73\times10^{-7}$
              & $2.11\times10^{6}$  \\
    \cline{4-9}
    &   &
        &  4  & resonant state & 97.8778 & 9.8933 & $3.32\times10^{-4}$
              & $3.01\times10^{3}$  \\
    \cline{4-9}
    &   &
        &  5  & resonant state & 112.3600 & 10.06 & $1.86\times10^{-2}$
              & $53.698$  \\
    \cline{4-9}
    &   &
        &  6  & resonant state & 121.6872 & 11.0312 & $0.2019$
              & $4.952$  \\
   \hline
\end{tabular}
\end{center}
\caption{The mass, width, and lifetime of bound or resonant KK modes of the fermions.
The parameters are set to $a=1$, $v=1$, $\lambda=10$, and $\eta=0, 5, 10$.} \label{tableLR2}
\end{table*}

\begin{table*}[h]
\begin{center}
\renewcommand\arraystretch{1.3}
\begin{tabular}
 {|l|c|c|c|c|c|c|c|c|c|}
 \hline
 $\lambda$  & $\eta$ & Height of $V_{\text{L}}$ & $n$  &
  Bound or resonant & $m^{2}$  &  $m$  &  $\Gamma$  & $\tau$    \\
    \hline\hline
 $15$  & $0$  & $V_{L}^{\textrm{PT}}(\infty)=33.3333$
        &  0   & bound state & 0 & 0 & 0 & $\infty$ \\
    \cline{4-9}
    &   &
        &  1   & bound state & 15.2586 & 3.9062 & 0 & $\infty$  \\
    \cline{4-9}
    &   &
        &  2   & bound state & 26.2153 & 5.1201 & 0 & $\infty$  \\
     \cline{2-9}
     & $5$   &  $V_{L}^{\text{max}}=86.2314$
        & 0    & bound state & 0 & 0 & 0 & $\infty$ \\
    \cline{4-9}
    &   &
        &  1  & bound state & 25.4736 & 5.0471 & 0  & $\infty$  \\
    \cline{4-9}
    &   &
        &  2  & resonant state & 47.1786 & 6.8687 & $7.42\times10^{-13}$
               & $1.35\times10^{12}$  \\
    \cline{4-9}
    &   &
        &  3  & resonant state & 64.9127 & 8.0568 & $4.38\times10^{-6}$
               & $2.284\times10^{5}$  \\
    \cline{4-9}
    &   &
        &  4  & resonant state & 78.2950 & 8.8485 & $4.52\times10^{-3}$
               & $221.25$  \\
    \cline{4-9}
    &   &
        &  5  & resonant state & 86.3584 & 9.2929 & $0.11159$
               & $8.96161$  \\

   \cline{2-9}
     & $10$   &  $V_{L}^{\text{max}}=165.24$
        & 0    & bound state & 0 & 0 & 0 & $\infty$ \\
    \cline{4-9}
    &   &
        &  1  & resonant state & 35.5690 & 5.9640 & $8.85\times10^{-16}$
              & $1.13\times10^{15}$  \\
    \cline{4-9}
    &   &
        &  2  & resonant state & 67.5881 & 8.2212 & $1.78\times10^{-15}$
              & $5.63\times10^{14}$  \\
    \cline{4-9}
    &   &
        &  3  & resonant state & 95.9463 & 9.7952 & $3.27\times10^{-11}$
              & $2.73\times10^{10}$  \\
    \cline{4-9}
    &   &
        &  4  & resonant state & 120.4832 & 10.9765 & $6.00\times10^{-7}$
              & $1.67\times10^{6}$  \\
    \cline{4-9}
    &   &
        &  5  & resonant state & 140.9340 & 11.8716 & $3.09\times10^{-4}$
              & $3.232\times10^3$  \\
    \cline{4-9}
    &   &
        &  6  & resonant state & 156.6930 & 12.5177 & $0.0162$
              & $61.55221$  \\
     \cline{4-9}
    &   &
        &  7  & resonant state & 166.8501 & 12.9171 & $0.1707$
              & $5.8578$  \\
   \hline
\end{tabular}
\end{center}
\caption{The mass, width, and lifetime of bound or resonant KK modes of the fermions.
The parameters are set to $a=1$, $v=1$, $\lambda=15$, and $\eta=0, 5, 10$.} \label{tableLR3}
\end{table*}

Next, we turn to investigate the effect of the coupling coefficient $\lambda$ on localization of the fermion KK mode. The left-- and right--chiral fermion KK modes and the corresponding resonant mass spectrum are solved and calculated for different values of $\lambda$. We only list the mass, width, and lifetime of the left--chiral fermion KK modes in Tables \ref{tableLR2} and \ref{tableLR3}, since the result of the right--chiral fermion KK modes is the same except the zero mode. From Tables \ref{tableLR1}, \ref{tableLR2}, and \ref{tableLR3}, it can be concluded that for the same coupling coefficient $\lambda$, with the increase of the coupling coefficient $\eta$, the number of fermion resonances also increases, and the number of fermion bound KK modes decreases. The reason is that,
for the same $\lambda$, as increasing $\eta$ the potential barriers at both sides of the origin of the extra dimension become higher, and the potential well of the left--chiral fermion KK modes at $z\approx0$ tends to deeper and narrower. The mass of the first massive KK mode also increases with $\eta$. On the other hand, for the potential of the right--chiral fermion KK modes, when $\eta\geq\frac{1}{27}av\lambda(4v^2\lambda-9\sqrt{3})$, we have $V_{R}(0)\geq V_{R}(\infty)$, and there is no right--chiral fermion KK bound state and no mass gap in the mass spctrum of the right--chiral fermion KK modes.

\begin{figure}[htb]
\begin{center}
\subfigure[$n=0$, bound state]{\label{fig_L0515}
\includegraphics[width=7cm]{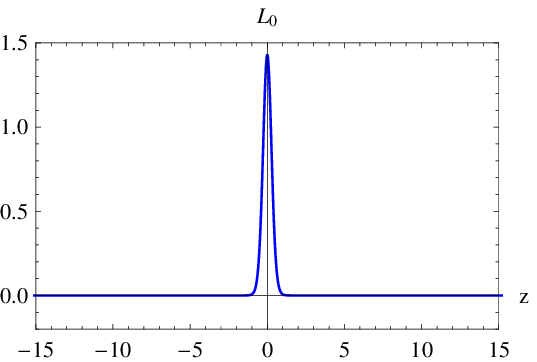}}
\subfigure[$n=1$, bound state]{\label{fig_L1515}
\includegraphics[width=7cm]{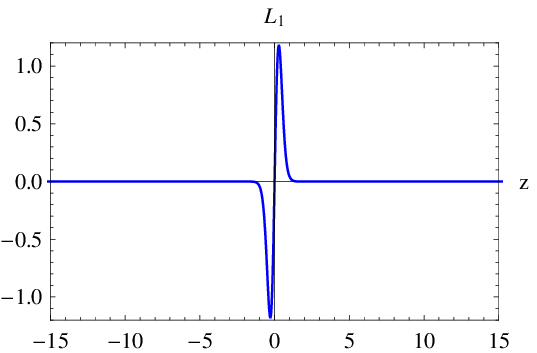}}
\subfigure[$n=2$, resonant state]{\label{fig_L2515}
\includegraphics[width=7cm]{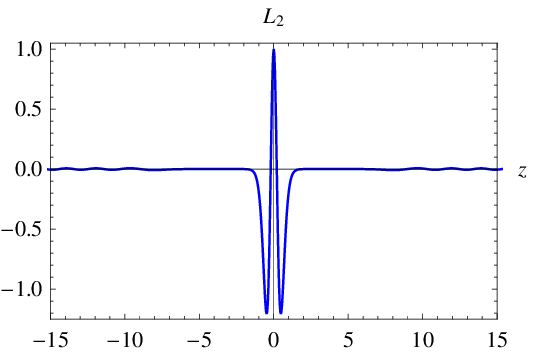}}
\subfigure[$n=3$, resonant state]{\label{fig_L3515}
\includegraphics[width=7cm]{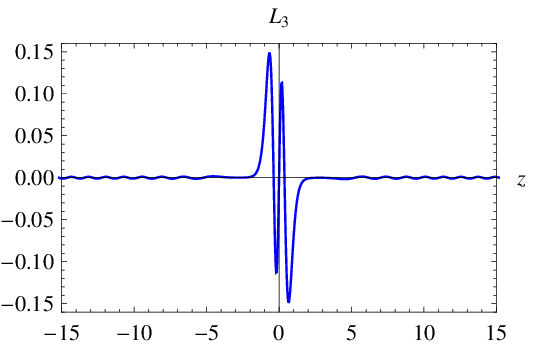}}
\subfigure[$n=4$, resonant state]{\label{fig_L4515}
\includegraphics[width=7cm]{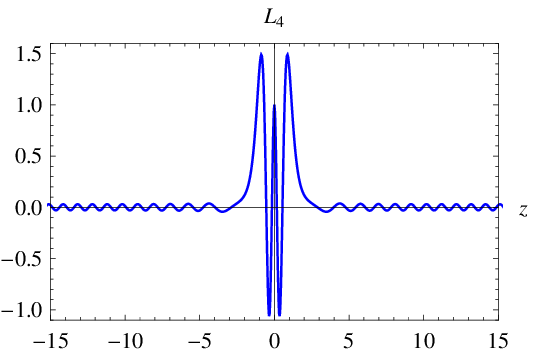}}
\subfigure[$n=5$, resonant state]{\label{fig_L5515}
\includegraphics[width=7cm]{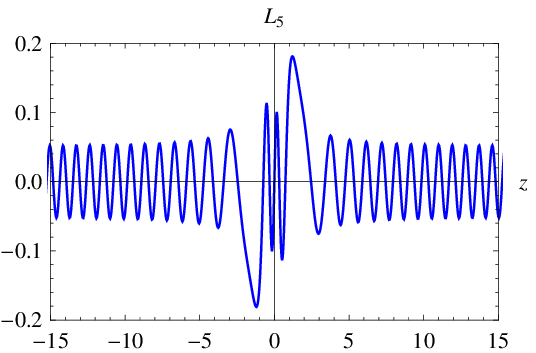}}
\end{center}\vskip -5mm
\caption{The shapes of the wave functions for the left--chiral fermion KK modes $L_n$, with the parameters $a=1$, $v=1$, $\lambda=15$, and $\eta=5$.}
 \label{fig_L515}
\end{figure}

\begin{figure}[htb]
\begin{center}
\subfigure[Left--chiral]{\label{fig_ml}
\includegraphics[width=7cm]{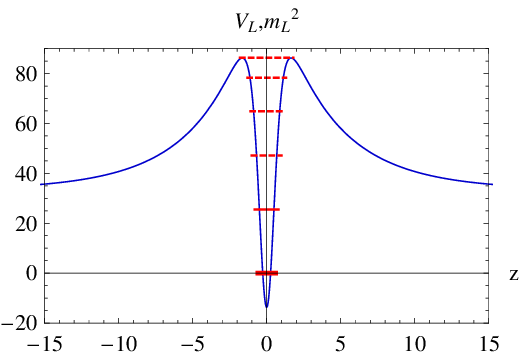}}
\subfigure[Right-chiral]{\label{fig_mr}
\includegraphics[width=7cm]{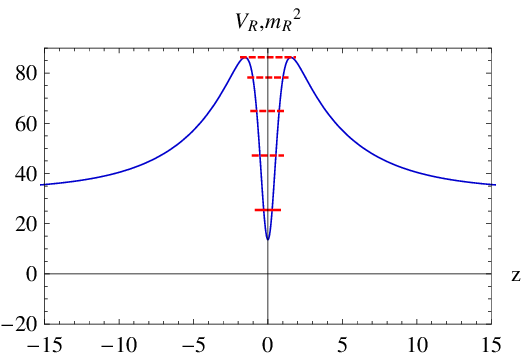}}
\end{center}\vskip -5mm
\caption{The mass spectra and effective potentials for the left-- and right--chiral fermion KK modes. The blue lines represent the potentials $V_{L}$ and $V_{R}$, the red lines represent the square of the mass for bound KK modes and the red dashed line represent the square of the mass for resonant KK modes. The parameters are set to $a=1$, $v=1$, $\lambda=15$, and $\eta=5$.}
 \label{fig_massspectra}
\end{figure}

For the same $\lambda$ and $\eta$, the lifetime of a resonant KK mode decreases with its mass. As an example, when $\lambda=15$ and $\eta=5$, all localization and quasilocalization of the left--chiral fermion KK modes are shown in Fig. \ref{fig_L515}. The zero mode and first massive KK mode are bound states, which are localized on the brane. Some other massive KK modes are resonances, which are quasilocalized on the brane. The mass spectra of the left-- and right--chiral fermion KK modes are also plotted in Fig. \ref{fig_massspectra}. Only the left--chiral fermion zero mode can be localized on the brane, so there exists only the four--dimensional massless left--chiral fermion. The first massive left-- and right--chiral fermion KK modes can also be localized on the brane, so a four--dimensional massive Dirac fermion consisting of a pair of coupled left-- and right--chiral KK modes can be localized on the brane. When $m_{L,R}^{2}\geq V_{L,R}(\infty)=\frac{4}{27}a^{2}v^{4}\lambda^{2}$, there are four pairs of coupled left-- and right--chiral resonant KK modes, so the four-dimensional massive Dirac fermions with finite lifetimes can be quasilocalized on the brane. When $m_{L,R}^{2}\gg V_{L,R}^{\textrm{max}}$, both the left-- and right--chiral fermion KK modes cannot be confined to the brane and will be excited into the bulk.

\section{conclusion and discussion}
\label{sec:conclusion}

In this paper, localization and mass spectra of a bulk spin-$1/2$ fermion field on a thick brane generated by two scalar fields (a kink scalar and a dilaton scalar) in five dimensions have been investigated.

For localization of the fermion zero mode, the coupling between the fermion and scalar fields must be introduced in the five--dimensional action. In this paper, we introduce two types of the couplings: the usual Yukawa coupling $-\eta\bar{\Psi}\phi\Psi$ between the fermion and kink scalar and the new coupling $\lambda\bar{\Psi}\Gamma^{M}(\partial_{M}\pi)\gamma^{5}\Psi$
between the fermion and dilaton scalar.

If only introducing the Yukawa coupling, both of the left-- and right--chiral fermion zero modes cannot be localized on the brane. So the new coupling term $\lambda\bar{\Psi}\Gamma^{M}(\partial_{M}\pi)\gamma^{5}\Psi$ must be introduced. If only introducing the new coupling, the left--chiral fermion zero mode is localized on the brane, and finite number of bound massive KK modes of left-- and right--chiral fermions may be localized on the brane. The number of bound states increases with the coupling coefficient $\lambda$. If we further introduce the Yukawa coupling, there exists a finite number of resonant massive KK modes of left-- and right--chiral fermions, and the number of resonances also increases with the Yukawa coupling coefficient $\eta$. Hence, the massless fermion localized on the brane is consisted of just the left--chiral KK mode, while the massive fermions localized or quasilocalized on the brane are consisted of the left-- and right-- chiral fermion KK modes and hence represent the four--dimensional Dirac massive fermions. The lifetime of the fermion KK resonant modes decreases with their masses.

\section*{Acknowledgement}

H. Guo was supported by the National Natural Science Foundation of China (Grants No. 11305119) and the Fundamental Research Funds for the Central Universities (Grants No. K5051307001 and No. K5051307019), Q.-Y. Xie was supported by the National Natural Science Foundation of China (Grants No. 11375075), and C.-E. Fu was supported by the National Natural Science Foundation of China (Grants No. 11405121).


\end{document}